\documentstyle[aps,preprint]{revtex}
\tightenlines
\begin{document}

\draft
\title{Quantitative analysis  by renormalized entropy
	of invasive electroencephalograph recordings in
	focal epilepsy}
\author{K.~Kopitzki}
\address{Center for Data Analysis and Modelling\\
         D-79 104 Freiburg, Germany\\ \& \\
         Dept. of Stereotactic Neurosurgery\\
         D-79 106 Freiburg, Germany}
\author{P.~C.~Warnke}
\address{Dept. of Stereotactic Neurosurgery\\
         D-79 106 Freiburg, Germany}
\author{J.~Timmer}
\address{Center for Data Analysis and Modelling\\
         D-79 104 Freiburg, Germany}
\date{February 16, 1998}
\maketitle

\begin{abstract}
Invasive electroencephalograph (EEG) recordings of ten 
patients suffering from focal epilepsy were analyzed
using the method of renormalized entropy. Introduced as
a complexity measure for the different regimes of a
dynamical system, the feature was tested here for its
spatio-temporal behavior in epileptic seizures. In all
patients a decrease of renormalized entropy within the
ictal phase of seizure was found. Furthermore, the strength
of this decrease is monotonically related to
the distance of the recording location to the focus.
The results suggest that the method of renormalized entropy is 
a useful procedure for clinical applications like
seizure detection and localization of epileptic foci.
\end{abstract}


\section{ Introduction}
Focal epilepsies are characterized by seizures which
originate from a distinct region of the brain.
The identification of this so called epileptogenic
focus is a prerequisite for surgical treatment.
Thus much attention is paid to the characterization of
invasive electroencephalograph (EEG) recordings of patients
suffering from this disease. Competitive techniques like 
high-resolution positron-emission tomography or MRI
using specific ligands or metabolites are used as
alternative to localize the epileptogenic focus in
order to avoid invasive EEG recordings. For the time
being no superiority of either method has been established
yet \cite{achten,weckess,semah,merlet}. Any method will be
preferred which has a better spatial and timewise
resolution than the other with acceptable validity.
Within this scenario a new approach in EEG-analysis
is described in this paper.\\
Since there is only little doubt about the nonlinearity
of the dynamical system underlying the observed time
series a broad range of nonlinear analysis techniques
has been applied to these data.
As one of the first results of nonlinear EEG analysis
indication of low-dimensional chaos was reported by
Babloyanz and Destexhe \cite{bab} who claimed a noninteger
correlation dimension in the EEG recording of a petit
mal seizure. Furthermore, Frank et al.\cite{frank} obtained
positive Lyapunov exponents by analyzing two seizure
recordings. Similar effects, giving indication of chaotic
behavior in EEG data, were also described by Freeman and
Skarda \cite{frsk}, Dvorak and Siska \cite{dvsk}, Basar
and Bullock \cite{babu} and Pijn et al. \cite{pijn}.
Windowed estimates of these measures for piecewise
classification of EEG were used by Lehnertz and Elger
\cite{lehn1}, Tirsch et al. \cite{tirsch}, Pritchard
et al. \cite{prit} and Iasemidis and Sackellares \cite{iasem}. 
Lehnertz et al. \cite{lehn1} analyzed the  EEG recordings
of 20 patients suffering form unilateral temporal lobe
epilepsy. The variability of the correlation dimension,
estimated for subsequent segments of the EEG, was found
to be a good indicator of the lateralisation of the
epileptic focus. In a case study Lerner \cite{lerner}
found the correlation integral itself to be suitable to
detect seizure activity in an EEG recording.\\  
By now it is well appreciated that results, obtained by
use of these measures, have to be interpreted with care.
Several investigators \cite{theiler92,rapp93,glass93}  
have especially shown that the value of the correlation
dimension is influenced by computational as well as
recording parameters. Theiler et al. \cite{theiler95}
found no evidence for low-dimensional chaos when re-examining
EEG data of 110 patients. The finite correlation dimensions,
they obtained  for the same data sets in an earlier study
\cite{rapp89}, were found to be caused by an artifact of
the autocorrelation in the oversampled signals.\\
A methodologically different approach is based on the examination
of neural spike trains which might be observed in the
EEG of epileptic patients. In a basic implementation the
statistical properties like mean, variance or skewness of the
interspike interval distribution are used to characterize the
EEG. Although this method neglects the sequence of these
intervals it is proven to be of practical use for some
applications \cite{mandell}. However, for analysis of
epileptic seizures advanced methods seemed to be necessary,
i.e sequence-sensitive methods \cite{dayh1,dayh2,legen,lesti}. 
Using a sequence-sensitive complexity measure Rapp et al. found
a decrease of nonrandom structure of these sequences in focal
epileptic seizures induced in rats by application of penicillin
\cite{rapp}.\\
A qualitative characterization of epileptic seizures is given
by Heyden et al.\cite{heyden}. Their results, obtained by
analyzing EEG data of patients suffering from mesial temporal
lobe epilepsy, indicate that the property of reversibility
of a time series can be used to discriminate between seizure
and non-seizure activity. Casdagli et al. \cite{casdagli}
report recurrent activity to occur in spatio-temporal 
patterns related to the location of an epileptic focus.\\
In the present paper a procedure is described which seems to
be appropriate to classify the EEG of epileptic patients in
an uniform way. The method of renormalized entropy, proposed
in \cite{saparin} for classification of the different
states of a dynamical system, is applied to the EEG data
of ten patients suffering from temporal lobe epilepsy.
The method is tested for its ability to assign a given segment
of an EEG time series to the corresponding neurophysiological
state of the epileptic patient, i.e.~interictal, ictal or
posictal phase, as well as its use for localization of an
epileptic focus.\\

\section{Renormalized Entropy}
Applying Klimontovich's S-theorem \cite{klim} to Fourier 
spectra of scalar time series, Saparin et al. \cite{saparin}
introduced renormalized entropy as a complexity measure for
the different regimes of a dynamical system. Here each regime
is represented by the normalized Fourier spectrum $S_{i}(\omega)$
of one observable $x_i(t)$ which, in formal analogy to classical
statistical physics, is viewed at as a distribution function of
an open system.\\
Given a reference distribution $S_{r}(\omega)$, representing
the systems state of equilibrium, the relative degree of order
of the regime described by $S_{i}(\omega)$ is determined by
comparing the entropies of these two distributions under the
additional constraint of same mean energy in both states.
To this end an effective Hamiltonian
\begin{equation}
H_{eff}=-\ln S_r(\omega)
\end{equation}
is introduced and the reference spectrum is renormalized
("heated up") into 
\begin{equation}
\tilde S_r(\omega)=C(T_i)e^{-\frac{H_{eff}(\omega)}{T_i}}=C(T_i) 
S_r(\omega)^{1/T_i},
\end{equation}
so that holds:
\begin{equation}
\int \tilde S_{r}(\omega) H_{eff}(\omega) d\omega=
\int S_{i}(\omega) H_{eff}(\omega) d\omega
\end{equation}
and
\begin{equation}
\int \tilde S_{r}(\omega) d\omega=1.
\end{equation}
Here eq.~(3) ensures the equality of mean energies of
the two states, eq.~(4) the normalization of the renormalized
spectrum. $C(T_i)$ is a normalization factor depending on $T_i$.\\ 
Renormalized entropy now is given by:
\begin{equation}
\Delta H=\int \tilde{S}_r(\omega) \ln \tilde{S}_r(\omega) d\omega-
\int S_i(\omega) \ln S_i(\omega) d\omega \quad .
\end{equation}
Applying this method to the different regimes of the
logistic map, Saparin et al. \cite{saparin} found
renormalized entropy to clearly detect all transitions
between the different types of periodic behavior as
well as the different types of chaos. Kurths et al.
\cite{kurt} and Voss et al. \cite{voss} analyzed
the heart rate variability of patients after myocardial
infarction. They report renormalized entropy to be
a suitable method for the detection of high risk
patients threatened by sudden cardiac death.\\
Because of eq.~(1) the method can only be applied to
processes which have a purely positive spectrum, i.~e.
chaotic or stochastic processes. Because spectra of
such processes have to be estimated the spectrum chosen
as reference should be of lower energy than each other
state of the system to avoid "temperatures" $T_i$ less
than 1. "Cooling down" an estimated spectrum increases
the variance of the estimate and therefore the variance of
the estimated entropy.

\section{Analysis of the EEG}
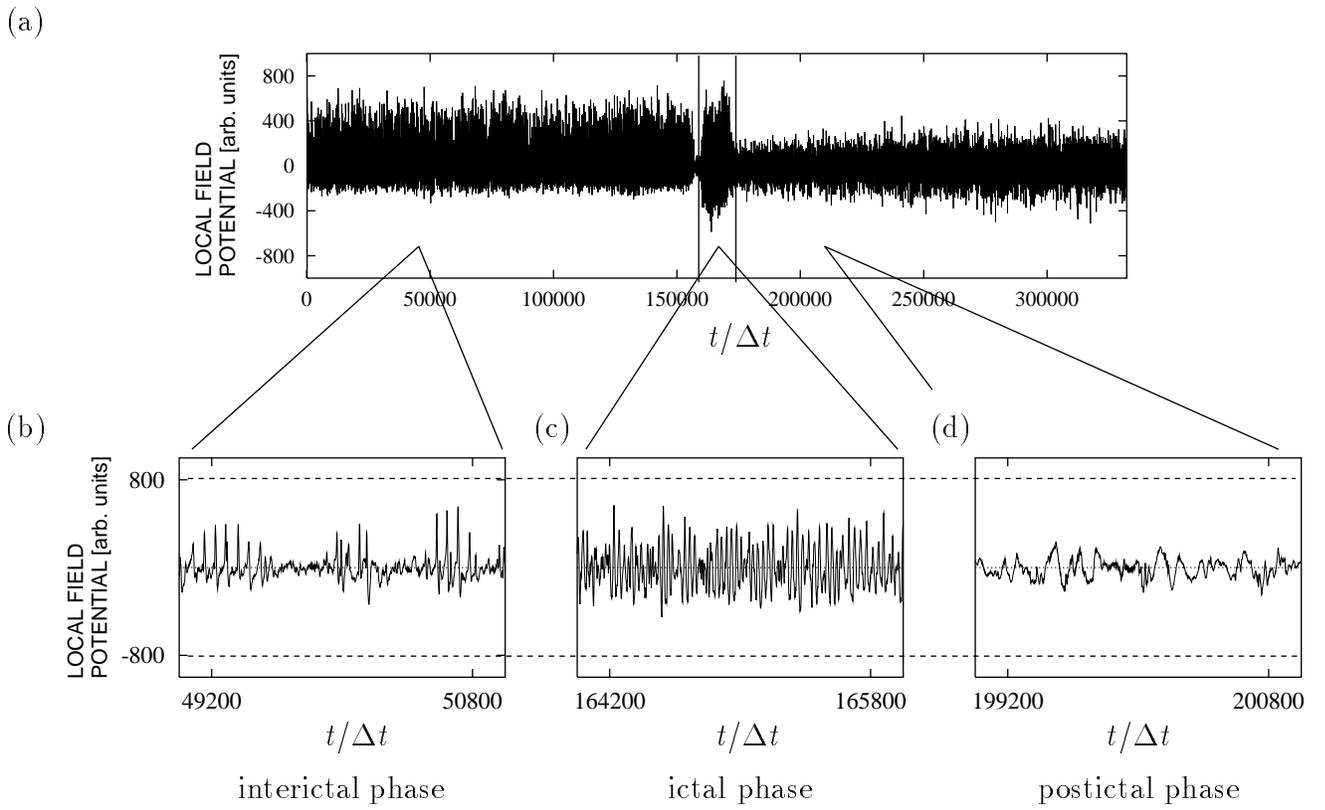
\begin{figure}
\caption{Invasive EGG recording of an epileptic seizure (a): segments
of the (b) interictal phase, (c) ictal phase and (d) postictal
phase. The vertical lines in (a) denote the beginning and the end
of the ictal phase. The measured local field potentials are shown
in arbitrary units.}
\label{epityp}
\end{figure}

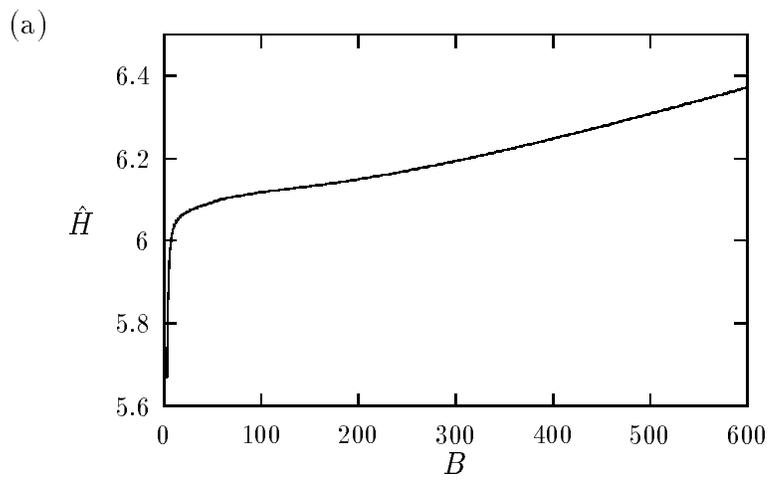
\begin{figure}
\caption{Estimate $\hat{H}$ of spectral entropy versus width $B$ of the
smoothing window obtained for a segment of an EEG (a) and an AR(2) process
(b) of length N=4096. The horizontal line in (b) denotes the true value
of spectral entropy.}
\label{ewidth}
\end{figure}

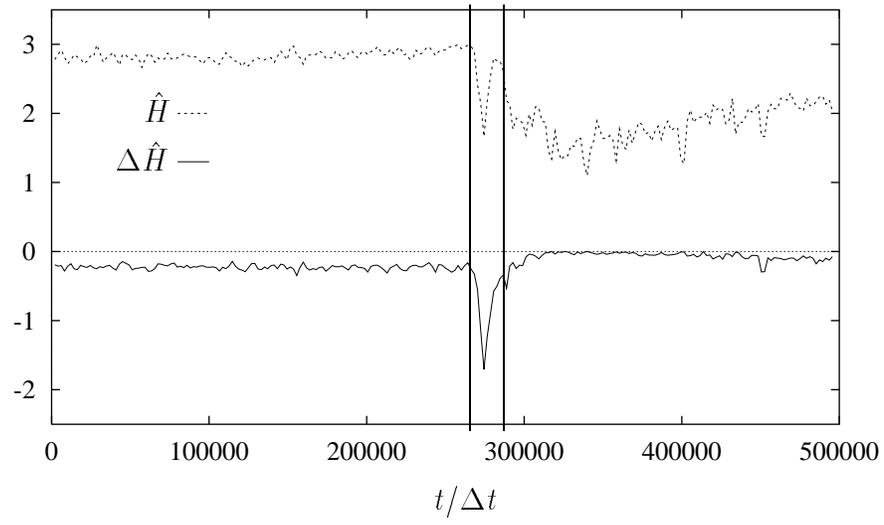
\begin{figure}
\caption{Course of spectral entropy $\hat{H}$ and renormalized entropy
$\Delta \hat{H}$ obtained for a recording location in the epileptogenic
area. Vertical lines denote the beginning and the end of the ictal phase.} 
\label{entr}
\end{figure}

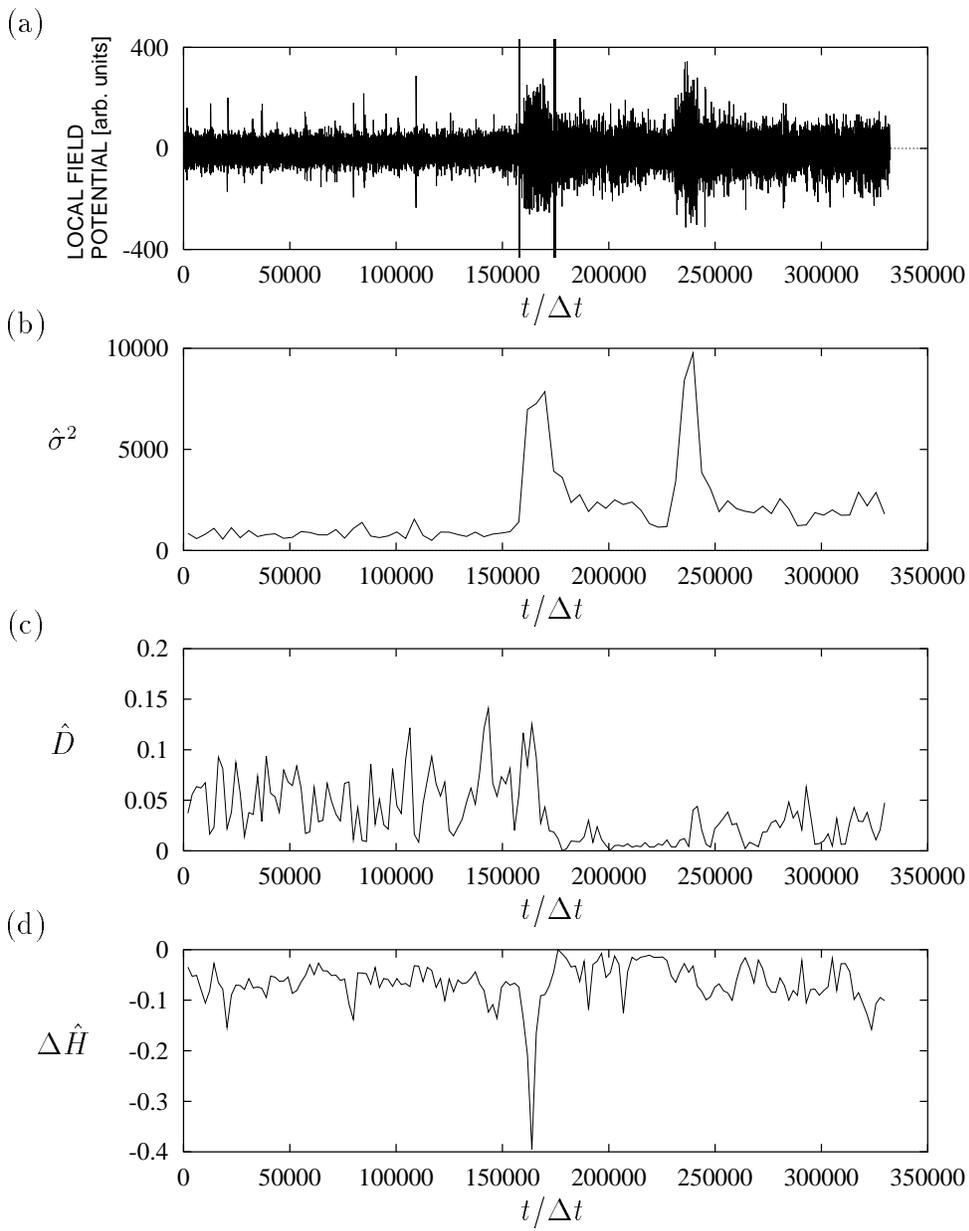
\begin{figure}
\caption{Course of (b) estimated variance $\hat{\sigma}^2$,
(c) euclidean distance $\hat{D}$ and (d) renormalized
entropy $\Delta \hat{H}$ obtained for the EEG shown in (a). Vertical lines
in (a) denote the beginning and the end of the ictal phase.} 
\label{comp}
\end{figure}

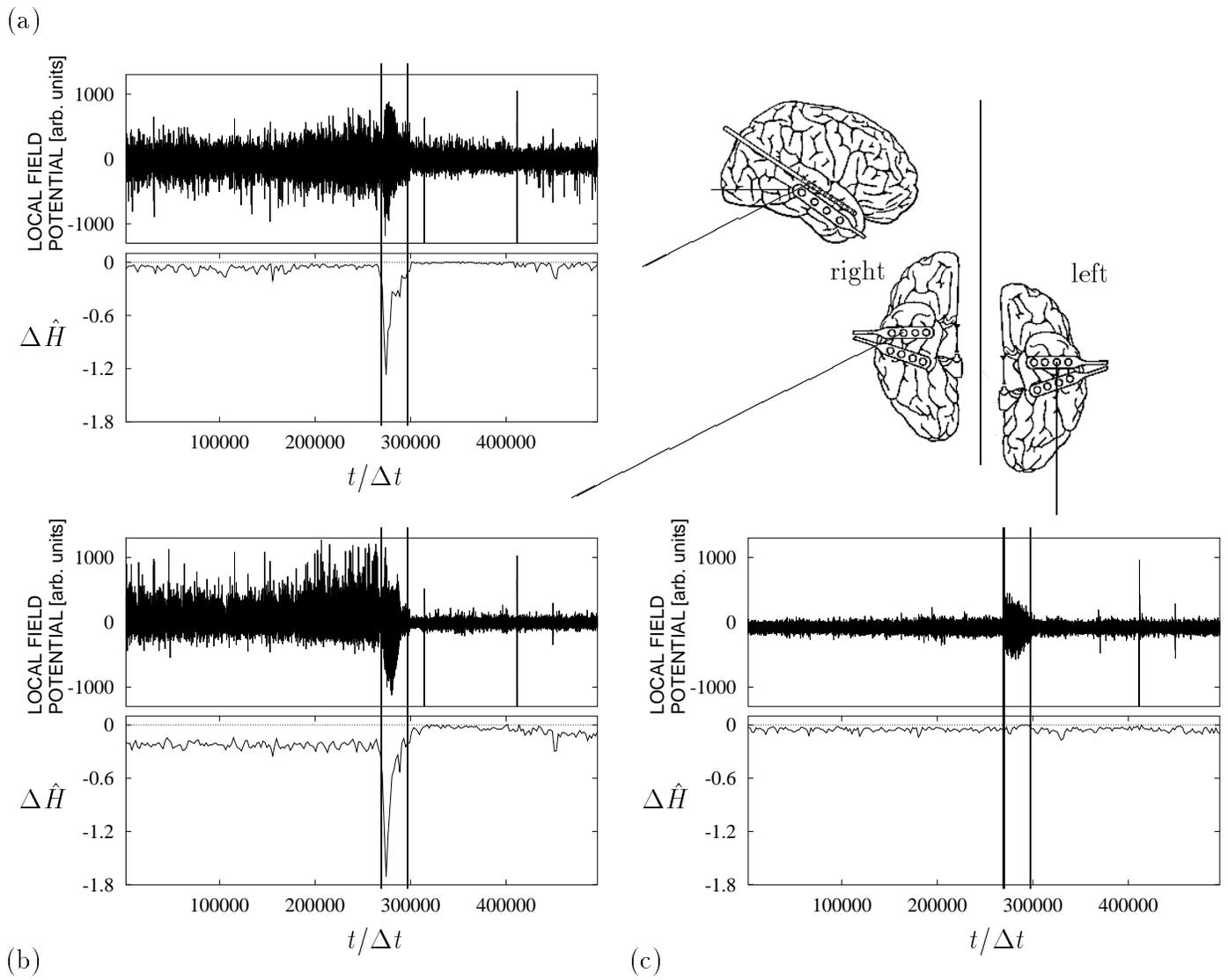
\begin{figure}
\caption{EEG and course of renormalized entropy obtained for recording
locations of different distance to the epileptic focus:
(a) location nearby the epileptic focus,
(b) location of smallest distance to the epileptic focus, (c) location
on the contralateral hemisphere. Vertical lines denote the beginning
and the end of the ictal phase.}
\label{ktanly}
\end{figure}

The EEG data analyzed in this study were recorded using
chronically implanted subdural and intrahippocampal
electrodes measuring the local field potential.The
obtained signals were passed to a multi-channel amplifier
system with band-pass filter settings of $0.53 Hz - 85 Hz$
and were written to a digital storage device with a
sampling interval of $\Delta t = 5.76\, ms$ per channel.
Fig.~\ref{epityp} displays representative samples of
the obtained time series for different phases of an
epileptic seizure. For each recording the identification of
the different phases was done by experienced clinicians
by visual inspection of the time series.\\
For analysis the data of each channel were divided into
consecutive segments $x_{i,.}$ of length $N=4096$ with
a $50\%$ overlap. The length corresponds to a duration
of 24 s per epoch and was chosen to achieve at least
quasi-stationarity for each segment according
to \cite{lopez}.\\
To apply the method described above to these data the
spectrum of each segment has to be estimated and for 
each channel a reference spectrum  has to be
found.\\
To estimate the spectra $S_{i}$ the periodograms
\begin{equation}
Per_i(\omega_k)=1/N\,|\sum_{j=1}^{N}x_{i,j}e^{i\omega_k j \Delta t}|^2
\end{equation}
were smoothed:
\begin{equation}
\hat{S}_i(\omega_k)=\sum_u w_u Per(\omega_{k-u}) \quad .
\end{equation}
As smoothing kernel $w_u$ the Bartlett-Priestley window
was chosen \cite{pries}:
\begin{equation}
w_{u}=\left\{\begin{array}{r@{\quad:\quad}l}
C\,(1-(\frac{u}{b})^{2})&|u|\leq b\\0&|u|>b \quad .
\end{array}\right.\
\end{equation}
Variance and bias of the estimator depend on the width
$B=2b+1$ of the smoothing window and the structure of
the true spectrum. To find an appropriate value for the
window width the spectral entropy
\begin{equation}
\hat{H}=-\sum_k \hat{S}(\omega_k) \ln \hat{S}(\omega_k)
\end{equation} 
was calculated as function of $B$ for different segments. 
Fig.\ref{ewidth}a shows the plot obtained for one segment
of an EEG. The graph can be divided into two regions. For
small values of $B$ the fluctuations of the periodogram
are suppressed insufficiently: because each summand in
eq.~(9) is a convex function of $S(\omega_k)$ the spectral
entropy is underestimated. In this region the estimated
spectral entropy increases fast with increasing $B$.
For large values of $B$ there is an area of small increase
where the periodogram is oversmoothed. Since information
about the structure of the spectrum is lost, in this region
the spectral entropy is overestimated.\\
Fig.\ref{ewidth}b shows the corresponding plot for one
realization of an AR(2) process
\begin{equation}
X(t)=a_{1}X(t-1)+a_2X(t-2)+\epsilon(t)
\end{equation}
with
\[a_1=1.3, a_2=-0.75 \quad \mbox{and} \quad \epsilon(t)\in
{\cal WN}(0,1).\]
of length $N=4096$.
This process describes a damped linear oscillator driven by white
noise \cite{honer}. The functional relationship between the
parameters $a_1$,$a_2$ and the frequency $\omega$ and relaxation
time $\tau$ of the oscillator is given by:
\begin{eqnarray}
a_{1}&=&2\cos\omega e^{-1/\tau}\\
a_{2}&=&-e^{-2/\tau}\qquad.
\end{eqnarray}
Because this process is linear its spectral entropy can be
calculated analytically:
\begin{equation}
H=-\sum_k S(\omega_k) \ln S(\omega_k)
\end{equation} 
with
\begin{equation}
S(\omega_{k})=
\frac{C}{|1-a_{1}e^{i\omega_{k}}-a_{2}e^{i2\omega_{k}}|^{2}}\quad ,
\end{equation}
$C$ a normalizing constant.
As the plot shows the true value of spectral entropy, denoted by
the horizontal line in Fig.\ref{ewidth}b, is reached in the area
of transition from high to low increase of spectral entropy.
Therefore, a value of $B=33$ from this region in Fig.\ref{ewidth}a
was chosen for the analysis of the EEG.\\
To calculate the renormalized entropy of the EEG spectra for each
channel a reference $S_r(\omega_k)$ has to be chosen. As mentioned
before this state should be of lower energy than each other state
of the system. Because
\begin{equation}
-\sum_k S_j(\omega_k) \ln S_r(\omega_k) \geq
-\sum_k S_r(\omega_k) \ln S_r(\omega_k)
\end{equation}
holds for every $j$ if
\begin{equation}
-\sum_k S_j(\omega_k) \ln S_j(\omega_k) \geq
-\sum_k S_r(\omega_k) \ln S_r(\omega_k)
\end{equation}
the spectrum of lowest spectral entropy (eq.~(9)) was chosen
as reference. If postictal phase of an epileptic seizure differed
from interictal phase the corresponding segment was found in the
beginning of the postictal phase resulting in a course of renormalized
entropy shown in Fig.\ref{entr}. Otherwise the reference was found in
the interictal phase. Fig.\ref{entr} also shows that the conventional
spectral entropy as given by eq.~(9) does not serve as a feature
characterizing the ictal phase.\\
To test and compare the behavior of the renormalized entropy in an
epileptic seizure, EEG data of all patients were analyzed using the
method of renormalized entropy as well as simple features like the
variance
\begin{equation}
\hat{\sigma}^2_i=\frac{1}{N-1}\sum_j(x_{i,j}-\frac{1}{N}\sum_l x_{i,l})^2
\end{equation}
or the squared euclidean distance
\begin{equation}
\hat{D}_i=\sum_k (\hat{S}_i(\omega_k)-\hat{S}_r(\omega_k))^2
\end{equation}
of spectra. For determination of the euclidean distance the
spectra were calculated in the same way as were done for
calculation of renormalized entropy. Also the distance was
calculated with respect to the same reference spectrum to
achieve results comparable to these obtained by use of renormalized
entropy. A representative sample is given
in Fig.\ref{comp}. In Fig.\ref{comp}a the EEG recording,
in Fig.\ref{comp}b the course of the estimated variance and in
Fig.\ref{comp}c the course of the squared euclidean distance obtained for
this recording are shown. By means of these simple
characteristics a reliable identification of the different phases
(interictal, ictal and postictal phase) of an epileptic seizure is
not possible. The squared euclidean distance which was chosen
as alternative and more elementary distance measure of spectra 
fails to distinguish between the different phases.
The variance detects the ictal phase but miss-classifies
a later postictal segment of the EEG. By way of contrast,
the course of renormalized entropy reveals a temporary strong
decrease only within the ictal phase.\\
To investigate the spatial behavior of renormalized entropy,
for each patient up to eight channels, corresponding to recording
locations of different distance to the epileptic focus, were analyzed.
In all patients the value of renormalized entropy within the ictal
phase was found to decrease with decreasing distance of the recording
location to the epileptic focus, as shown for a representative example in
Fig.\ref{ktanly}. Thus, a technical device for localizing epileptic foci,
based on the concept of renormaized entropy, is imaginable.

\section{Conclusions}
The method of renormalized entropy, formally introduced to quantify the
complexity of the different regimes of a dynamical system, has been
applied to invasive EEG recordings of ten patients suffering from
temporal lobe epilepsy.\\
In all patients the course of renormalized entropy obtained for recording
locations nearby the epileptic focus shows a strong decrease in the ictal
phase of an epileptic seizure with respect to the interictal or postictal
phase. Because the strength of this decrease depends on the distance of
the recording location to the focus not only a discrimination between the
different phases but also a localization of the focus seems to be possible.\\
The method makes exclusively use of the spectral properties of the time
series under consideration and therefore human interaction is restricted
to the choice of the spectral estimator to be used.\\
Putting it altogether the concept of renormalized entropy seems to be a
promising candidate for clinical applications like seizure detection or
localization of epiletic foci.

\section{Acknowledgment}
The data, analyzed in this study, were kindly made available by
C.~E.~Elger and K.~Lehnertz from the University Clinic of
Epileptology, Bonn. We would like to thank P.~David for initiating
this study and P. Saparin for critically reading the manuscript.

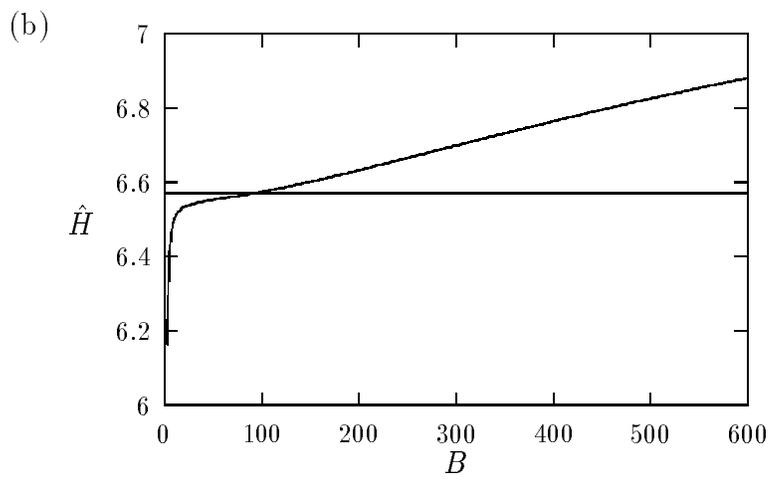
\end{document}